**Suppression of the Superconducting Transition in $R$FeAsO$_{1-x}$F$_x$ for $R$ = Tb, Dy, Ho**

Jennifer A. Rodgers,[1,2] George B. S. Penny,[1,2] Andrea Marcinkova,[1,2] Jan-Willem G. Bos,[1,2] Dmitry A. Sokolov,[1,3] Anna Kusmartseva,[1,2] Andrew D. Huxley [1,3] and J. Paul Attfield [1,2]*

[1]*Centre for Science at Extreme Conditions, University of Edinburgh, King's Buildings, Mayfield Road, Edinburgh, EH9 3JZ.*
[2]*School of Chemistry, University of Edinburgh, Edinburgh, EH9 3JJ.*
[3]*SUPA, School of Physics, University of Edinburgh, Edinburgh, EH9 3JZ.*

A suppression of superconductivity in the late rare earth $R$FeAsO$_{1-x}$F$_x$ materials is reported. The maximum critical temperature (T$_c$) decreases from 51 K for $R$ = Tb to 36 K for HoFeAsO$_{0.9}$F$_{0.1}$, which has been synthesised under 10 GPa pressure. This suppression is driven by a decrease in the Fe-As-Fe angle below an optimimum value of 110.6°, as the angle decreases linearly with unit cell volume (V) across the $R$FeAsO$_{1-x}$F$_x$ series. A crossover in electronic structure around this optimum geometry is evidenced by a change in sign of the compositional dT$_c$/dV, from negative values for previously reported large $R$ materials to positive for HoFeAsO$_{0.9}$F$_{0.1}$.

Rare earth ($R$) oxypnictides $R$FeAsO[1] were recently discovered to superconduct when doped, with critical temperatures surpassed only by the high-T$_c$ cuprates. Several families of superconducting iron pnictides have subsequently been discovered.[2] These all have layered structures containing AsFeAs slabs with Fe tetrahedrally coordinated by As. The main types are the 1111 materials based on $R$FeAsO or $M$FeAsF ($M$ = Ca, Sr, Ba), the 122 phases $M$Fe$_2$As$_2$, and the 111 $A$FeAs ($A$ = Li, Na) family. The related binaries FeX (X = Se, Te) are also superconducting.

The electron-doped 1111 materials $R$FeAsO$_{1-x}$F$_x$ and $R$FeAsO$_{1-\delta}$ materials remain prominent as they have the highest T$_c$'s, up to 56 K, and allow lattice and doping effects to be investigated through variations of the $R^{3+}$ cation size and the anion composition. A strong lattice effect is evident at the start of the rare earth series, as T$_c$ rises from 26 K for LaFeAsO$_{1-x}$F$_x$ to 43 K under pressure,[3,4] and to a near-constant maximum 50-56 K in the $R$FeAsO$_{1-x}$F$_x$ and $R$FeAsO$_{1-\delta}$ series for $R$ = Pr to Gd,[5,6,7,8,9,10] but whether lattice effects ultimately enhance or suppress superconductivity for the late $R$'s has been unclear. The late rare earth $R$FeAsO$_{1-x}$F$_x$ materials and the oxygen-deficient $R$FeAsO$_{1-\delta}$ superconductors require high pressure synthesis, leading to significant challenges as single phase samples are difficult to prepare, and accurate analyses of cation stoichiometries and O and F contents are difficult. To investigate the effect of the lattice for later $R$ we have synthesised multiple samples of $R$FeAsO$_{0.9}$F$_{0.1}$ ($R$ = Tb, Dy, and Ho) under varying high pressure conditions. Here we report superconductivity in HoFeAsO$_{0.9}$F$_{0.1}$ for which the maximum T$_c$ of 36 K is markedly lower than in the previous $R$ analogs. This is part of a systematic suppression of superconductivity by the smaller, late $R$ cations. HoFeAsO$_{0.9}$F$_{0.1}$ also shows a reversal in the sign of the compositional dT$_c$/dV (V = unit cell volume) compared to the early $R$ materials, confirming that the decreasing $R$ size has a significant effect on the bands contributing to the Fermi surface.

Polycrystalline ceramic $R$FeAsO$_{1-x}$F$_x$ samples ($R$ = Tb, Dy, and Ho) were synthesised by a high pressure method and investigated by powder X-ray diffraction, magnetization and conductivity measurements.[11] Initial results for $R$FeAsO$_{1-x}$F$_x$ ($R$ = Tb and Dy) were published elsewhere.[12] Both materials were found to be superconducting with maximum $T_c$'s of 46 and 45 K respectively. Little difference in superconducting properties between samples with nominal compositions of x = 0.1 and 0.2 were observed, and the x = 0.2 materials were generally of lower phase purity, and so the x = 0.1 composition was used in subsequent syntheses. The best samples typically contain ~80% by mass of the superconducting phase with residual non-superconducting $R_2O_3$ and $R$As phases also present. The sample purity and superconducting properties are not sensitive to synthesis pressure over a range that moves to higher pressures as $R$ decreases in size; $R$ = Tb and Dy superconductors were respectively prepared at 7-10 and 8-12 GPa, heating at 1050-1100 °C. Repeated syntheses of TbFeAsO$_{1-x}$F$_x$ gave several samples with higher $T_c$'s than the above value, the highest value is $T_c$(max) = 51 K (Fig. 1). Further DyFeAsO$_{1-x}$F$_x$ samples did not show higher transitions than before, so we conclude that $T_c$(max) in this system is 45 K.

Tetragonal HoFeAsO$_{0.9}$F$_{0.1}$ was obtained from reactions at 10 GPa pressure and the properties of six HoFeAsO$_{0.9}$F$_{0.1}$ samples prepared under varying conditions are summarized in Table 1. Crystal structure refinements and phase analysis were carried out by fitting powder X-ray diffraction data (Fig. 2).[13] Magnetisation measurements demonstrate that all six HoFeAsO$_{1-x}$F$_x$ samples are bulk superconductors with $T_c$'s of 29-36 K (Fig. 3). Resistivities show smooth high temperature evolutions without apparent spin density wave anomalies. The transitions to the zero resistance state have widths of less than 4 K.

Although all of the samples in Table 1 have the same starting composition, small variations of synthesis pressure and temperature result in a dispersion in x around the nominal 0.1 value for the HoFeAsO$_{1-x}$F$_x$ phase and corresponding variations in superconducting properties. T$_c$ increases to a maximum value, T$_c$(max), at the upper solubility limit of x in $R$FeAsO$_{1-x}$F$_x$ systems,[7] and this is consistent with the observation that the superconducting phases in samples 1, 3 and 4, which are heated at high temperatures or longer times and so are likely to have a slightly lower F content, have lower T$_c$'s (average 32.1 K) than the other three samples, made under nominally identical 'optimum' conditions, which have average T$_c$ = 34.8 K. Sample 6 shows the highest T$_c$ = 36.2 K and the lowest proportion of the HoFeAsO$_{1-x}$F$_x$ phase and a correspondingly low diamagnetic volume fraction. This demonstrates that the sample is at the upper limit of the superconducting composition range and so gives a realistic T$_c$(max) for the HoFeAsO$_{1-x}$F$_x$ system.

Although the doping values x for the high pressure $R$FeAsO$_{1-x}$F$_x$ samples are not known precisely, comparing ensembles of samples with similar phase purities made under similar conditions reveals a clear suppression of superconductivity by lattice effects for heavier $R$. For example, all of our TbFeAsO$_{1-x}$F$_x$ superconductors have higher T$_c$'s (five TbFeAsO$_{1-x}$F$_x$ samples, T$_c$ = 45-51 K) than all of the HoFeAsO$_{1-x}$F$_x$ materials (in Table 1). The T$_c$(max) values of 51, 45 and 36 K for $R$FeAsO$_{1-x}$F$_x$ with $R$ = Tb, Dy and Ho, respectively thus represent the trend correctly.

Fig. 4 shows a plot of the maximum critical temperatures, $T_c$(max), against unit cell volume for many reported $R$FeAsO$_{1-x}$F$_x$ and $R$FeAsO$_{1-\delta}$ systems and our above materials. $T_c$(max) rises slowly as cell volume decreases for R = La to Pr and then shows a broad maximum, between $R$ = Pr and Tb in the $R$FeAsO$_{1-x}$F$_x$ materials, before falling rapidly as $R$ changes from Tb to Dy to Ho. This trend is not seen in the reported $R$FeAsO$_{1-x}$ superconductors, where $T_c$(max) remains approximately constant,[14,15] apparently because they have larger cell volumes than their $R$FeAsO$_{1-x}$F$_x$ analogs (see Fig. 4).

The size of the $R^{3+}$ cation tunes the electronic properties through variations in the geometry of the FeAs slab. A trend between the As-Fe-As (or equivalent Fe-As-Fe) angle and $T_c$ has been reported for the early $R$ materials.[16] The upper panel of Fig. 4 shows representative reported values for optimal $R$FeAsO$_{1-x}$F$_x$ superconductors including our $R$ = Tb, Dy, and Ho materials. This demonstrates that the angle decreases monotonically with $R$ size and so does not show a universal correlation with $T_c$(max). The $T_c$(max) variation in the $R$FeAsO$_{1-x}$F$_x$ series is described by a simple $\cos(\phi-\phi_0)$ function, shown in Fig. 4, where the value of the As-Fe-As angle corresponding to the global maximum $T_c$, $\phi_{max}$ = 110.6°, is close to the ideal 109.5° value for a regular FeAs$_4$ tetrahedron. All five of the Fe 3d-bands are partially occupied and contribute to the Fermi surface of the iron arsenide superconductors through hybridization with As 4s and 4p states.[17] Decreasing the tetrahedral angle through 109.5° marks the crossover from tetragonal compression to elongation of the FeAs$_4$ tetrahedra. In a crystal field model, this reverses the splittings of the $t_2$ and e d-orbital sets and so a significant crossover in the real electronic structure is likely to occur near 109.5°.

Evidence for the above crossover also comes from a discovered change in the sign of the compositional $dT_c/dV$ near optimum doping in the $R$FeAsO$_{1-x}$F$_x$ systems.[18] The unit cell parameters and volume for the six HoFeAsO$_{1-x}$F$_x$ samples in Table 1 show a positive correlation with $T_c$ (Fig. 5), in contrast to early $R$ = La[19] and Sm[7] analogs where lattice parameters and volume decease with increasing $T_c$. The $T_c$,V points for near-optimally doped $R$ = La, Sm and Ho $R$FeAsO$_{1-x}$F$_x$ superconductors are shown on Fig. 4 together with the derived $dT_c/dV$ values. $dT_c/dV$ for a single $R$FeAsO$_{1-x}$F$_x$ system follows the overall trend in $dT_c(max)/dV$ for different $R$'s, changing from a negative value at large $R$ = La to a small positive slope at $R$ = Ho.

The compositional $dT_c/dV$ for a given $R$FeAsO$_{1-x}$F$_x$ system reflects two competing effects of variations in the fluoride content x on the lattice volume. $F^-$ is slightly smaller than $O^{2-}$ so the anion substitution effect gives a negative contribution to the compositional $dT_c/dV$, independent of $R$. The concomitant effect of doping electrons into the Fe d-bands tends to expand the lattice (and increase $T_c$), but the magnitude of this positive $dT_c/dV$ term depends on the nature of the bands at the Fermi surface. The observed shift from negative to positive $dT_c/dV$ as $R$ changes from La to Ho shows that the decreasing size of the $R^{3+}$ cation leads to significant changes in the Fermi surface, with volume-expanding (antibonding) bands more prominent for smaller $R$. Calculations have confirmed that the electronic structure near the Fermi level is sensitive to such small changes in the As z-coordinate (equivalent to changing the Fe-As-Fe angle).[20] Small changes in the contributions of the d-bands are likely to be particularly important in a multigap scenario for superconductivity, as evidenced in gap measurements of TbFeAsO$_{0.9}$F$_{0.1}$ and other iron arsenide materials.[21]

In summary, our analysis of multiple samples of $R$FeAsO$_{1-x}$F$_x$ ($R$ = Tb, Dy, and Ho) superconductors demonstrates that the maximum critical temperature falls from 51 K for $R$ = Tb to 36 K for the previously unreported Ho analog. Hence, the effect on the lattice of substituting smaller, late rare earths in the $R$FeAsO$_{1-x}$F$_x$ lattice suppresses superconductivity. This lattice control appears to be through tuning of the interatomic angles in the FeAs layer, with the optimum angle being 110.6°, near the ideal tetrahedral value. The compositional dT$_c$/dV changes sign around the optimum angle evidencing significant changes in the Fermi surface. It appears difficult to increase the critical temperatures above 56 K in 1111 type iron arsenide materials through tuning lattice effects, although the possibility of higher T$_c$'s in other structure types remains open.

We acknowledge EPSRC, the Royal Society of Edinburgh and the Leverhulme trust for support.

* Corresponding author: j.p.attfield@ed.ac.uk

**Table 1:** Synthesis conditions (all samples were synthesised at 10 GPa), refined lattice parameters and volume, $T_c$'s, mass fractions and superconducting volume fractions for HoFeAsO$_{1-x}$F$_x$ samples.

| Sample | $t_{synth}$ (hr) | $T_{synth}$ (°C) | a (Å) | c (Å) | Vol (Å$^3$) | $T_c$ (K) | Mass frac. (%) | Diamag. frac. (%) |
|---|---|---|---|---|---|---|---|---|
| 1 | 2 | 1150 | 3.8246(3) | 8.254(1) | 120.74(3) | 29.3 | 75 | 70 |
| 2 | 2 | 1100 | 3.8272(2) | 8.2649(8) | 121.06(2) | 33.0 | 74 | 85 |
| 3 | 1 | 1150 | 3.8258(5) | 8.264(2) | 120.96(4) | 33.2 | 73 | 76 |
| 4 | 3 | 1100 | 3.8282(5) | 8.261(2) | 121.07(5) | 33.7 | 84 | 74 |
| 5 | 2 | 1100 | 3.8282(2) | 8.2654(7) | 121.13(2) | 35.2 | 81 | 57 |
| 6 | 2 | 1100 | 3.8297(7) | 8.270(2) | 121.30(7) | 36.2 | 58 | 46 |

**Fig. 1** Resistivity and (inset) susceptibility data for an optimum sample of TbFeAsO$_{0.9}$F$_{0.1}$, showing a sharp superconducting transition at $T_c$ = 51 K. The sample was prepared at 7 GPa and 1050 °C.

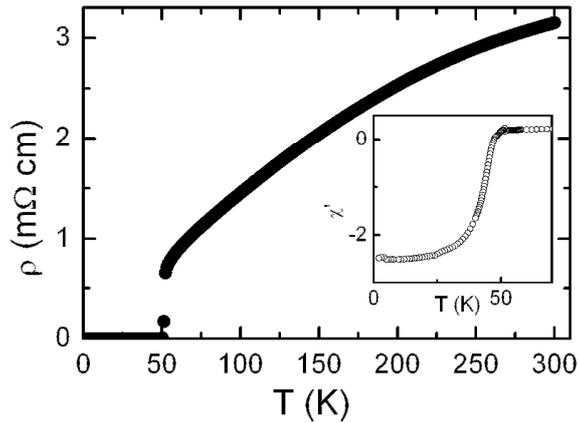

**Fig. 2** Fitted x-ray diffraction profile for HoFeAsO$_{0.9}$F$_{0.1}$ (sample 5) at room temperature. The Bragg markers (from top to bottom) are for the minority phases, Ho$_2$O$_3$ and HoAs, and for HoFeAsO$_{0.9}$F$_{0.1}$.

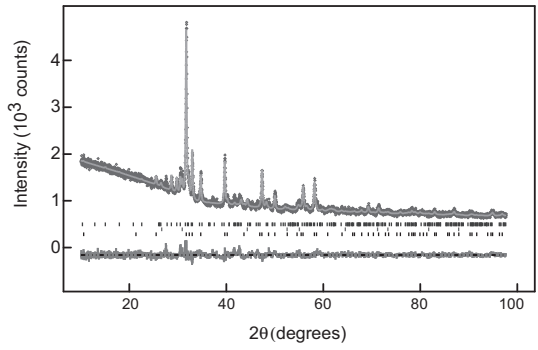

**Fig. 3** Superconductivity measurements for HoFeAs$_{0.9}$F$_{0.1}$; (a) ac magnetic volume susceptibility for the six samples; (b) resistivities for samples 4 and 6.

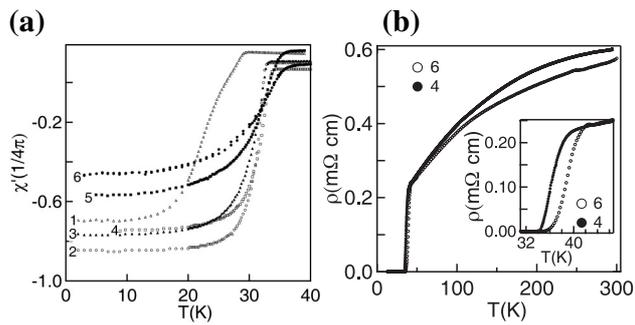

**Fig. 4** Variation of Fe-As-Fe angle $\phi$ (upper panel) and superconducting $T_c$ (lower panel) with unit cell volume for different $RFeAsO_{1-x}F_x$ (circles)[19,22,5,7,12] and $RFeAsO_{1-\delta}$ (triangles)[14,15]. $T_c(max)$ points are shown as filled symbols. The fit of equation $T_c(max) = T_c(max)_0 \cdot \cos A(\phi - \phi_0)$ with parameters $T_c(max)_0 = 56$ K, $A = 0.03$, and $\phi_0 = 110.6°$ is also shown. $dT_c/dV$ values are derived from the data for sub-optimally doped materials (open symbols) in the $R$ = La,[19] Sm[7] and Ho (this paper) systems.

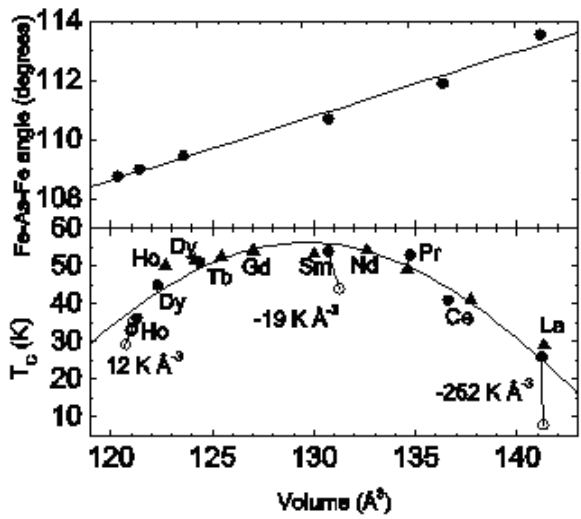

**Fig. 5** Variations of $T_c$ with the tetragonal unit cell parameters and volume for the six $HoFeAsO_{1-x}F_x$ samples in Table 1.

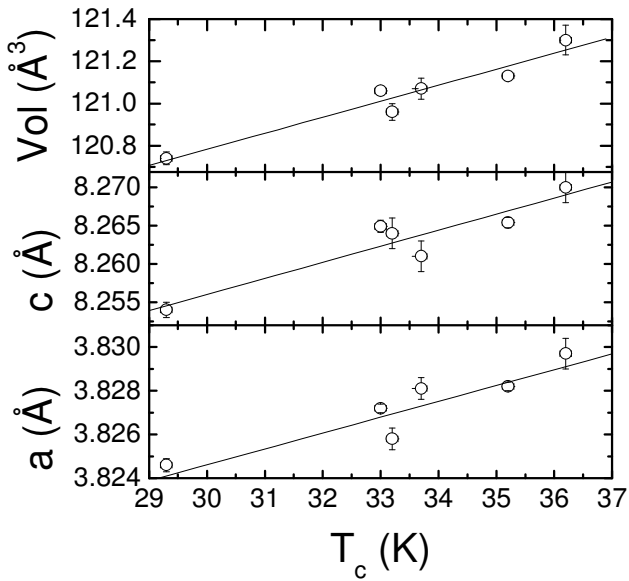

[11] Samples were synthesised from stoichiometric amounts of $R$As, $Fe_2O_3$, $FeF_2$ and Fe, using a Walker multianvil module within a 1000 tonne press. The products were dense, black, sintered polycrystalline pellets. Powder X-ray diffraction data were collected on a Bruker AXS D8 diffractometer using Cu K$\alpha_1$ radiation. Data were recorded at $10 \leq 2\theta \leq 100°$ with a step size of $0.007°$ for Rietveld analysis. ac magnetic susceptibility was

measured from 3 to 50 K with a field of 0.5 Oe oscillating at 117 Hz using a Quantum Design SQUID magnetometer. Electrical resistivity was measured by a four-probe method between 1.7 and 300 K using a Quantum Design physical property measurement system and an APD cryogenics closed cycle refrigeration unit with an in-house built sample stage.

[12] J.-W. G. Bos, G. B. S. Penny, J. A. Rodgers, D. A. Sokolov, A. D. Huxley and J. P. Attfield, *Chem. Comm.* **31,** 3634 (2008).

[13] $HoFeAsO_{0.9}F_{0.1}$ has a tetragonal structure (space group P4/nmm; results from fit shown in Fig. 2; goodness of fit $\chi^2 = 1.60$, residuals; $R_{wp} = 3.94\%$, $R_p = 3.02\%$; cell parameters $a = 3.8282(2)$ Å, $c = 8.2654(7)$ Å; atom positions (x,y,z) and isotropic temperature (U) factors; Ho (¼,¼,0.1454(4)), 0.044(2) Å$^2$; As (¼,¼,0.6659(5)), 0.029(2) Å$^2$; Fe (¾,¼,½), 0.014(2) Å$^2$; O,F (¾,¼,0), 0.26(2) Å$^2$). The secondary $Ho_2O_3$ phase is in a high pressure B-type rare earth oxide modification, space group *C2/m*, $a = 13.841(2)$ Å, $b = 3.4984(5)$ Å, $c = 8.608(1)$ Å, $\beta = 100.08(1)°$.

[14] K. Miyazawa, K. Kihou, P. M. Shirage, C. H. Lee, H. Kito, H. Eisaki, and A. Iyo, *J. Phys. Soc. Jpn.* **78,** 034712 (2009).

[15] J. Yang, X. L. Shen, W. Lu, W. Yi, Z. C. Li, Z. A. Ren, G. C. Che, X. L. Dong, L. L. Sun, F. Zhou, and Z. X. Zhao, *New J. Phys.* **11**, 025005 (2009).

[16] J. Zhao, Q. Huang, C. de la Cruz, S. Li, J. W. Lynn, Y. Chen, M. A. Green, G. F. Chen, G. Li, Z. Li, J. L. Luo, N. L. Wang and P. Dai, *Nature Mater.* **7**, 953 (2008).

[17] D. J. Singh and M. –H. Du, *Phys. Rev. Lett.* **100**, 237003 (2008).

[18] The compositional $dT_c/dV$ quantifies the changes in $T_c$ and unit cell volume V due to variations in doping level x at constant (atmospheric) pressure, and is complementary to

the pressure-induced $dT_c/dV$ at constant x. Both derivatives are negative for LaFeAsO$_{1-x}$F$_x$, and we thus predict a positive pressure-induced $dT_c/dV$ (pressure suppression of superconductivity) for HoFeAsO$_{1-x}$F$_x$.